\documentclass[aps,preprint,onecolumn,secnumarabic,nobalancelastpage,amsmath,amssymb,
nofootinbib]{revtex4}

\RequirePackage{fix-cm}
\usepackage{enumerate}

\usepackage{graphicx}      
\usepackage{longtable}     
\usepackage{url}           
\usepackage{dcolumn}
\usepackage{hyperref}

\usepackage{bm}            
\usepackage{mathrsfs}
\usepackage{epstopdf}
\usepackage{color}
\begin{document}

\preprint{APS/123-QED}

\title{A study of geodesic motion in a (2+1)--dimensional charged BTZ black hole}

\author{Saheb Soroushfar}
\author{Reza Saffari}
\email{rsk@guilan.ac.ir}
\author{Afsaneh Jafari}

\affiliation{Department of Physics, University of Guilan,
41335-1914, Rasht, Iran.}

\date{\today}

\begin{abstract}
This study is purposed to derive the equation of motion for
geodesics in vicinity of spacetime of a $(2+1)$--dimensional charged
BTZ black hole. In this paper, we solve geodesics for both massive
and massless particles in terms of Weierstrass elliptic and Kleinian
sigma hyper--elliptic functions. Then we determine different
trajectories of motion for particles in terms of conserved energy
and angular momentum and also using effective potential.
\end{abstract}

\maketitle

\section{INTRODUCTION}

Black hole is one of the most interesting predictions of general
theory of relativity which has been attractive for theoretical
physicists for a long time, and it has still unknown parts to study.

Black hole is a region of spacetime with a strong gravitational
field that even light can not escape from it. It has an event
horizon which its total area never decreases in any physical
process \cite{Singh:2014gva}. In 1992 Banados, Teitelboim, and
Zanelli (BTZ) demonstrated that there is a black hole solution to
(2+1)--dimensional general relativity with a negative cosmological
constant \cite{Banados:1992wn} which it is proved that this type of
black hole arises from collapsing matter \cite{Ross:1992ba}. In
their
solution of gravitational field equation, it is required a constant
curvature in local spacetime \cite{Horowitz:1993jc}, which was a
strange result as a solution of general relativity. In a certain
subset of Anti--de Sitter (AdS) spacetime, they found a solution
which contains all the properties of black hole, by making a special
identification \cite{Horowitz:1993jc},\cite{Banados:1992gq}. Also,
the charged BTZ black hole is the analogous solution of AdS--Maxwell
gravity in (2+1)--dimension
\cite{Martinez:1999qi},\cite{Carlip:1995qv},\cite{Clement:1995zt}.

The BTZ black hole is interesting because of its connections with
string theory \cite{Sfetsos:1997xs},\cite{Hyun:1997jv} and its role
in microscopic entropy derivations
\cite{Carlip:1994gy},\cite{Carlip:1996yb}. The BTZ black hole can
also be used in some ways to study black holes in quantum
scales \cite{Carlip:1995qv},\cite{Strominger:1996sh}. Against the
Schwarzschild and Kerr black holes the BTZ black hole is
asymptotically anti--de Sitter rather than flat which has not
curvature singularity at the origin \cite{Carlip:1995qv}.

Black holes have various aspects to study. One of them that we are
more interested to investigate is the gravitational effects on test
particles and light which reach to spacetime of a black hole. It is
important because the motion of matter and light can be used to
classify an arbitrary spacetime, in order to discover its
structure. For this purpose, we need to solve geodesic equations
that describe the motion of particles and light. The analytical
solutions for many famous spacetimes (such as Schwarzschild
\cite{Y.Hagihara:1931}, four-dimensional Schwarzschild-de-Sitter
\cite{Hackmann:2008zz}, higher--dimensional Schwarzschild,
Schwarzschild--(anti)de Sitter, Reissner--Nordstrom and
Reissner--Nordstrom--(anti)-de Sitter \cite{Hackmann:2008tu}, Kerr
\cite{Kerr:1963ud}, Kerr--de Sitter \cite{Hackmann:2010zz}, A black
hole in f(R) gravity \cite{Soroushfar:2015wqa}) have been found
previously. The solutions are given in terms of Weierstrass
$\wp$-functions and derivatives of Kleinian sigma functions.


The interesting classical and quantum properties of the black hole, have made it appropriate to 
have existing a lower dimensional analogue that could represent the main features without 
unessential complications~\cite{Banados:1992wn}. Moreover, (2+1)-dimensional black holes 
are interesting as simplifed models for analyzing conceptual issues such as black hole 
thermodynamics~\cite{Ashtekar:2002qc}. In addition, the study of black holes in lower 
dimensions is useful to better understanding the physical features (like entropy, radiated flux) 
in a black hole geometry~\cite{Sa:1995vs}. Also, studying the gravitational field of (2+1)-dimensional 
black holes and motion around these black hole, can be useful.

The purpose in this paper is to determine types of particle's motion
around a (2+1)--dimensional charged BTZ black hole by studying its
spacetime. The outline of our paper is as follows. In section
\ref{ME} we introduce the metric and obtain geodesic equations.
Section \ref{AS} includes analytical solutions for massless and
massive particles and also the resulting orbits are classified in terms of
the energy and the angular momentum of test particle, and we conclude in section \ref{con}.
%

\section{Metric and geodesic equations} \label{ME}

The charged BTZ black hole is the solution of the (2+1)--dimensional
Einstein-Maxwell gravity with a negative cosmological constant
$\Lambda =-\frac{1}{l^2}$ \cite{Martinez:1999qi}. In the case of a
special matter source which is a nonlinear electrodynamic term in
the form of $(F_{\mu\nu}~F^{\mu\nu})^s$, which is called
Einstein-PMI gravity \cite{Hassaine:2008pw, Maeda:2008ha,
Hendi:2009zzc}, the form of the coupled (2+1)--dimensional action in
presence of cosmological constant is written as
follow \cite{Hendi:2010px}
\begin{equation}\label{I}
I(g_{\mu\nu}, A_\mu)=\frac{1}{16\pi}{\int}_{\partial M} d^3
x\sqrt{-g}[R-2\Lambda +(kF)^s].
\end{equation}
Here $R$ denotes the scalar curvature, $F$ is the Maxwell invariant which is equal to $F_{\mu\nu} F^{\mu\nu}$ ($F_{\mu\nu} = \partial_\mu A_\nu - \partial _\nu A_ \mu$ is the
electromagnetic tensor field and $A_\mu$ is the gauge potential),
and $s$ is an arbitrary positive nonlinearity parameter
($s\neq\frac{1}{2}$). Varying the action (\ref{I}) with respect to
$g_{\mu\nu}$ (the metric tensor) and $A_{\mu}$ (the electromagnetic
field), one can obtain the equations of gravitational and
electromagnetic fields as
\begin{equation}\label{G}
G_{\mu\nu} - \Lambda g_{\mu\nu} = T_{\mu\nu},
\end{equation}
\begin{equation}
\partial_\mu (\sqrt{-g}F^{\mu\nu}(kF)^{s-1})=0,
\end{equation}
and energy--momentum tensor is
\begin{equation} \label{Tmn}
T_{\mu\nu}=2[skF_{\mu\rho} F_\nu ^\rho (kF)^{s-1} - \frac{1}{4}g_{\mu\nu}(kF)^s],
\end{equation}
where
$k$ is a constant. When $s$ and $k$ go to $-1$, Eqs.(\ref{I}-\ref{Tmn}), reduce to the field equations of black hole in Einstein-Maxwell gravity. It is convenient to restrict the nonlinearity
parameter to $s>\frac{1}{2}$ in order to have asymptotically
well-defined electric field. The metric of non rotating charged BTZ
black hole can be written as following \cite{Hendi:2014mba}
\begin{equation}
ds^2 = -g(r)dt^2 + \frac{dr^2}{g(r)} +r^2 d\phi^2,
\end{equation}
in which the metric function $g(r)$ using the components
of Eq.(\ref{G}) obtains as \cite{Hendi:2014mba}
\begin{equation} \label{gr}
g(r)=\frac{r^2}{l^2} - m + \left\{
\begin{array}{rl}
{\ 2q^2 ln (\frac{r}{l})\ \ \ } \ \ \ \ \ \ \ \ \ \ \ \ \qquad \qquad   &s=1 ,\\\\
\frac{(2s-1)^2(\frac{sq^2(s-1)^2}{(2s-1)^2})^s}{2(s-1)}r^{(\frac{2(s-1)}{2s-1})} \ \ \ \ \  & otherwise.\\
\end{array} \right.
\end{equation}

This spacetime is characterized by $m$ (an integration constant
related to the mass), $q$ (the electric charge of the black hole)
and cosmological constant $\Lambda$. In the
case of $s=\frac{3}{4}$, one can obtain a well-known
metric which is called conformally invariant Maxwell solution \cite{Hendi:2014mba}, such as
\begin{equation}
g(r)=\frac{r^2}{l^2}-m-\frac{(2q^2)^{\frac{3}{4}}}{2r}.
\end{equation}

Taking $(2q^2)^{\frac{3}{4}}=K $  we have
\begin{equation} \label{metric}
ds^2= -(\frac{r^2}{l^2}-m-\frac{K}{2r})dt^2 +
\frac{dr^2}{\frac{r^2}{l^2} -m -\frac{K}{2r}} +r^2 d\phi^2.
\end{equation}

The metric (\ref{metric}) is stationay and axially symmetric. To
describe geodesic motion in such a spacetime we need geodesic
equation which is written as
\begin{equation}
\frac{d^2 x^{\mu}}{d\lambda^2}+\Gamma_{\rho\sigma}^\mu
\frac{dx^\rho}{d\lambda} \frac{dx^\sigma}{d\lambda}=0,
\end{equation}
in which $d\lambda^2=g_{\mu\nu} dx^\mu dx^\nu$ is the proper time
and  $\Gamma_{\rho\sigma}^\mu$  denotes the Christoffel connections
given by
\begin{equation}
\Gamma^\mu_{\rho \sigma} =\dfrac{1}{2} g^{\mu \nu}(\partial
_{\rho}{g_{\sigma \nu}}+\partial_{\sigma}{g_{\rho
\nu}}-\partial_{\nu} g_{\rho \sigma}).
\end{equation}

We can obtain geodesic equations using Lagrangian equation
\begin{equation} \label{L}
L=\frac{1}{2}\sum_{\mu,\nu=0}^3 g_{\mu\nu}\frac{dx^\mu}{d\lambda}
\frac{dx^\nu}{d\lambda} =\frac{1}{2} \epsilon
=\frac{1}{2}[-(\frac{r^2}{l^2}-m-\frac{K}{2r})(\frac{dt}{d\lambda})^2
+\frac{1}{(\frac{r^2}{l^2}-m-\frac{K}{2r})}(\frac{dr}{d\lambda})^2+r^2
(\frac{d\phi}{d\lambda})^2],
\end{equation}
where $\epsilon$ for massive and massless particles has the value of
$1$ and $0$ respectively and $\lambda$ is an affine parameter.

Using Euler--Lagrange equation we can obtain constants of motion
\begin{equation} \label{constants}
P_t=\frac{\partial L}{\partial \dot{t}}=
-(\frac{r^2}{l^2}-m-\frac{K}{2r})\dot{t}= -E \,\,\, ,\,\,\, P_\phi=
\frac{\partial L}{\partial \dot{\phi}}=r^2\dot{\phi}=\mathcal{L},
\end{equation}
where $E$ is energy and $\mathcal{L}$ is angular momentum. Now,
using Eq.(\ref{L}) and Eq.(\ref{constants}), we can obtain geodesic
equations as following
\begin{equation} \label{moadele}
(\frac{dr}{d\lambda})^2=E^2 +m\epsilon - \frac{\mathcal{L}^2}{l^2} -
\frac{\epsilon r^2}{l^2} + \frac{K\epsilon}{2r} + \frac{m
\mathcal{L}^2}{r^2} +\frac{K \mathcal{L}^2}{2r^3},
\end{equation}
\begin{equation} \label{phii}
(\frac{dr}{d\phi})^2 = (-\frac{\epsilon}{l^2 \mathcal{L}^2})r^6 +
(\frac{E^2}{\mathcal{L}^2}+\frac{m\epsilon}{\mathcal{L}^2}
-\frac{1}{l^2})r^4 +(\frac{K\epsilon}{2\mathcal{L}^2})r^3 +mr^2
+\frac{Kr}{2}=R(r),
\end{equation}
\begin{equation}
(\frac{dr}{dt})^2 =(\frac{r^2}{l^2}-m-\frac{K}{2r})^2 -
\frac{\epsilon (\frac{r^2}{l^2}-m-\frac{K}{2r})^3}{E^2} -
\frac{\mathcal{L}^2 (\frac{r^2}{l^2}-m-\frac{K}{2r})^3}{E^2 r^2}.
\end{equation}
These equations give a complete description of dynamics. Using
Eq.(\ref{moadele}) we can find effective potential
\begin{equation}
V_{eff}=\frac{\epsilon r^2}{l^2}-\frac{K\epsilon}{2r}
-\frac{m\mathcal{L}^2}{r^2} - \frac{K \mathcal{L}^2}{2r^3}
-m\epsilon +\frac{\mathcal{L}^2}{l^2}.
\end{equation}
Here for convenience we define a series of dimensionless parameters
as
\begin{equation}
\tilde{r}=\frac{r}{m} \,\,,\,\, \tilde{l}=\frac{l}{m}\,\,  ,\,\,
\tilde{K}=\frac{K}{m} \,\, ,\,\,
\tilde{\mathcal{L}}=\frac{m^2}{\mathcal{L}^2},
\end{equation}
and then rewrite Eq.(\ref{phii}) as
\begin{equation} \label{tilde}
(\frac{d\tilde{r}}{d\phi})^2 =
{\tilde{r}}^6(\frac{-\epsilon\tilde{\mathcal{L}}}{\tilde{l}^2}) +
{\tilde{r}}^4(E^2 \tilde{\mathcal{L}} + \epsilon \tilde{\mathcal{L}}
m -\frac{1}{\tilde{l}^2}) + \tilde{r}^3(\frac{\epsilon
\tilde{\mathcal{L}} \tilde{K}}{2})  + m \tilde{r}^2 +\frac{\tilde{K}
\tilde{r}}{2} = R(\tilde{r}).
\end{equation}
\subsection*{Comparision to other cases of paremeter $s$}
For $s=\frac{3}{2}$, the metric function is equal to $g(r)= \frac{r^2}{l^2} - m + A r^{\frac{1}{2}}$, in which $A=4 (\frac{3 q^2}{32})^{\frac{3}{2}}$, so we have
\begin{equation}
(\frac{dr}{d\phi})^2= r^6 (-\frac{\epsilon}{\mathcal{L}^2 l^2}) + r^4 (\frac{E^2}{\mathcal{L}^2}+ \frac{m\epsilon}{l^2 \mathcal{L}^2 } - \frac{1}{l^2}) + r^2(-\frac{A\epsilon}{\mathcal{L}^2} + m) - Ar.
\end{equation}
The solution of this equation is similar to Eq.(\ref{phii}) (i.e. for $s=\frac{3}{4}$, that, it is investigated completely in this paper).\\
In the case of $s=1$, the metric function $g(r)$ is
\begin{equation}
g(r)= \frac{r^2}{l^2} - m + 2 q^2 \ln (\frac{r}{l}),
\end{equation}
and so we have 
\begin{equation} \label{ln}
(\frac{dr}{d\phi})^2 = r^6(-\frac{\epsilon}{\mathcal{L}^2 l^2}) + r^4(\frac{E^2}{\mathcal{L}^2} + \frac{m\epsilon}{\mathcal{L}^2} - \frac{2 \epsilon q^2}{\mathcal{L}^2} \ln (\frac{r}{l}) - \frac{1}{l^2}) + r^2(m - 2q^2 \ln (\frac{r}{l})),
\end{equation}
and for $s=3$, the metric function is $g(r)= \frac{r^2}{l^2} - m + Q r^{\frac{4}{5}}$, where $Q= \frac{25 (\frac{12 q^2}{25})^3}{4}$, so we have
\begin{equation} \label{Q}
(\frac{dr}{d\phi})^2= r^6(- \frac{\epsilon}{\mathcal{L}^2 l^2}) + r^4 (\frac{E^2}{\mathcal{L}^2} + \frac{m\epsilon}{\mathcal{L}^2} - \frac{1}{l^2}) + mr^2 - \frac{Q \epsilon}{\mathcal{L}^2} r^{\frac{24}{5}} - Q r^{\frac{14}{5}}.
\end{equation}
Equation (\ref{ln}) includes some logarithmic terms, equation (\ref{Q}) and other equations related to other cases of $s$, have some terms with fractional powers of $r$, that, in our knowledge can not be solved analytically. However, they may be solved numerically similar to applied methods in Ref.~\cite{Hartmann:2010rr}. Therefore, in the following we consider the conformally invariant Maxwell solution ($s=\frac{3}{4}$).

\subsection*{Possible regions for Geodesic motion}
Equation (\ref{tilde}) implies that a necessary condition for the
existence of a geodesic is $R(\tilde{r})\ge0$, and therefore,
the real positive zeros of $R(\tilde{r})$ are extremal values of the
geodesic motion and determine the type of geodesic. Since
$\tilde{r}=0$ is a zero of this polynomial for all values of the
parameters, we can neglect it. So the Eq.(\ref{tilde}) changes to a
polynomial of degree $5$ as below
\begin{equation}
R^*(\tilde{r})
={\tilde{r}}^5(\frac{-\epsilon\tilde{\mathcal{L}}}{\tilde{l}^2}) +
{\tilde{r}}^3(E^2 \tilde{\mathcal{L}} + \epsilon \tilde{\mathcal{L}}
m -\frac{1}{\tilde{l}^2}) + \tilde{r}^2 (\frac{\epsilon
\tilde{\mathcal{L}} \tilde{K}}{2} ) + m\tilde{r}
+\frac{\tilde{K}}{2}.
\end{equation}

Using analytical solutions, one can analyze possible orbits which
depend on the parameters of test particle or light ray $\epsilon$,
$E^2$, $l$, $K$ and $\mathcal{L}$. In the next sections it will be
shown exactly.

For a given set of parameters $\epsilon$, $l$, $E^2$, $K$ and
$\mathcal{L}$ the polynomial $R^*(r)$ has a certain number of
positive and real zeros. If $E^2$ and $\mathcal{L}$ are varied, the
number of zeros can change only if two zeros of $R^*(r)$ merge to one.
Solving $R^*(\tilde{r}) =0$ and $\frac{d R^*(\tilde{r})}{d\tilde{r}}
=0$ give us $E^2$ and $\mathcal{\tilde{L}}$. For massive
particles $(\epsilon =1)$ we have
\begin{equation}
\tilde{\mathcal{L}}=-\frac{\tilde{l}^2 (4m\tilde{r} +
3\tilde{K})}{\tilde{r} ^2 (\tilde{K} \tilde{l} ^2 + 4 \tilde{r}
^3)}\,\,\,,\,\,\, E^2= - \frac{4\tilde{l}^4 m^2 \tilde{r}^2 +
4\tilde{K} \tilde{l}^4 m \tilde{r} -8\tilde{l}^2 m \tilde{r}^4
+\tilde{K}^2 \tilde{l}^4 - 4\tilde{K} \tilde{l}^2 \tilde{r}^3 + 4
\tilde{r}^6}{(4m\tilde{r} + 3\tilde{K}) \tilde{l}^4 \tilde{r}},
\end{equation}
and for massless particles $(\epsilon =0)$
\begin{equation}
\tilde{\mathcal{L}}=(-\frac{64 m^3}{27
\tilde{K}^2}+\frac{1}{\tilde{l}^2})\frac{1}{E^2}.
\end{equation}

\begin{figure}[ht]
\centerline{\includegraphics[width=7.25cm]{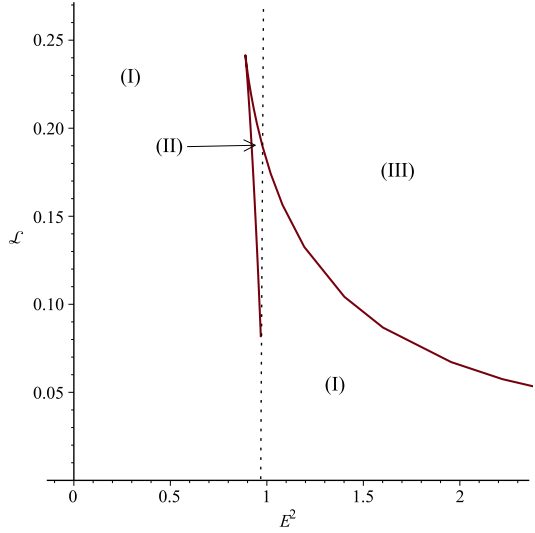}}
\caption{\label{massivespacetime}\small Region of different types of
geodesic motion for test particles ($\epsilon=1$). The numbers of
positive real zeros in these regions are: I=2, II=4, III=0. }
\end{figure}

\begin{figure}[ht]
\centerline{\includegraphics[width=7.25cm]{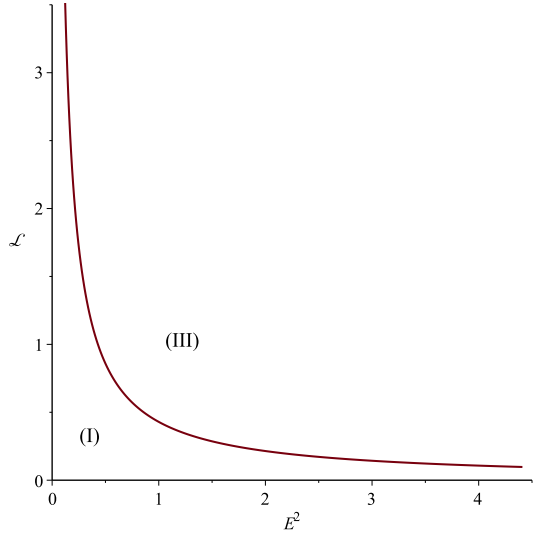}}
\caption{\label{masslessspacetime}\small  Region of different types
of geodesic motion for light ($\epsilon=0$). The numbers of positive
real zeros in these regions are: I=2, III=0.}
\end{figure}

The results of this analysis are shown in
Figs.(\ref{massivespacetime}), (\ref{masslessspacetime}) in which
regions of different types of geodesic motion are classified. \clearpage

The shape of an orbit is related to energy and angular momentum of
test particle. Since $\tilde{r}$ must be real and positive, the
acceptable physical regions can be found with the condition $E^2
\geqslant V_{eff}$. So the number of positive and real zeros of
$R(\tilde{r})$ will characterize the shape of different orbits. Here
according to the obtained results in this section, we
can identify three regions for geodesic motion of test particles:

1. In region I, $R^*(\tilde{r})$ has two real and positive zeros
$(r_1<r_2)$ which for $R^*(\tilde{r})\geq 0$ we have
$0<\tilde{r}<r_1$ and $\tilde{r}\geq r_2$. There are two kinds of
orbits, terminating bound orbit and flyby orbit (TBOs, FOs).\\

2. In region II, $R^*(\tilde{r})$ has four real positive zeros
$(r_i < r_{i+1})$ that for $R^*(\tilde{r})\geq 0$ they are
$0<\tilde{r}<r_1$, $r_2 <\tilde{r} <r_3$ and $r_4 \leq \tilde{r}$.
Three possible orbits are terminating bound, bound and flyby orbits
respectively (TBOs, BOs, FOs).\\

3. In region III, there is no real and positive zero for
$R^*(\tilde{r})$ and $R^*(\tilde{r})\geq 0$ for positive
$\tilde{r}$, therefore there is just terminating escape
orbit (TEOs).\\

\begin{figure}[ht]
\centerline{\includegraphics[width=7.25cm]{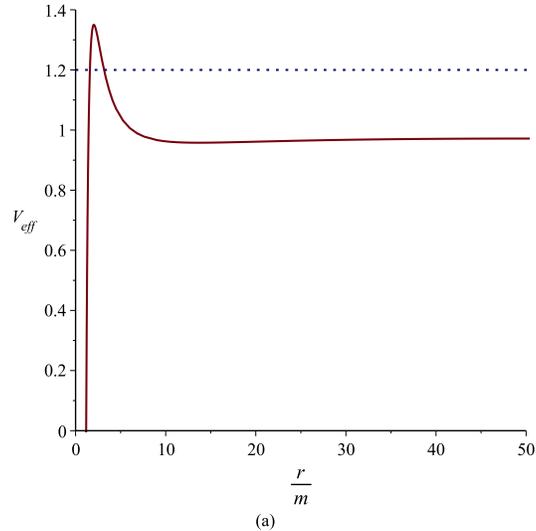}}
\centerline{\includegraphics[width=7.25cm]{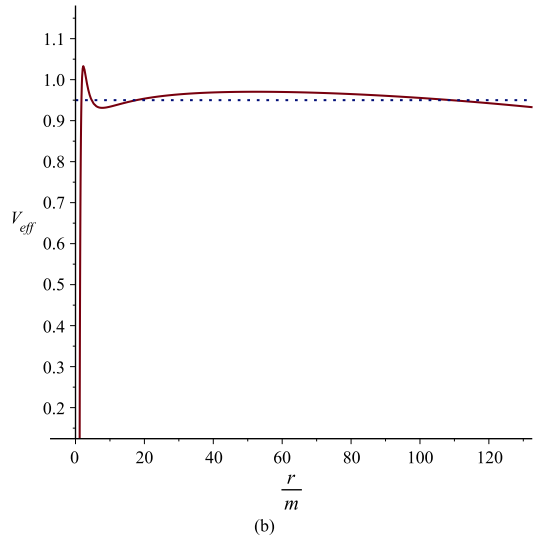}}
\centerline{\includegraphics[width=7.25cm]{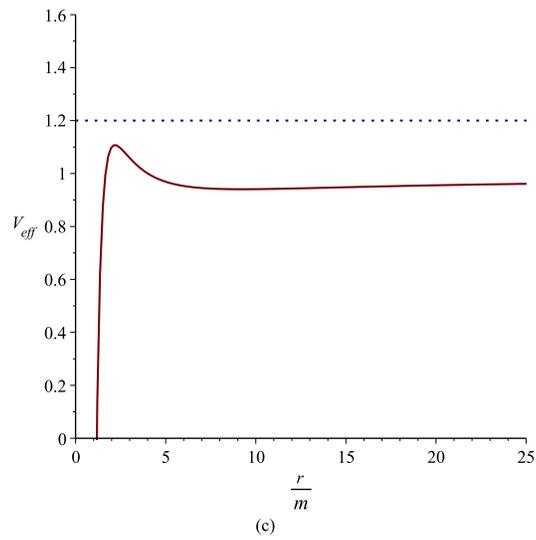}}
\caption{\label{potentials}\small Effective potentials for different
regions of geodesic motion for test particles (a: region I, b:
region II, c: region III). The horizontal line denotes the squared
energy parameter $ E^{2} $.}
\end{figure}

For timelike geodesics these three regions will appear but for null
geodesics only regions I and III are exist. In
Fig.\ref{potentials} different potentials for each of these
regions are illustrated.

\clearpage

\section{Analytical solution of geodesic equation} \label{AS}

In this section we study analytical solutions of equations of
motion. Using a new parameter $u=\frac{1}{\tilde{r}}$ we simplify
Eq.(\ref{tilde}) to
\begin{equation} \label{du}
(\frac{du}{d\phi})^2=\frac{\tilde{K}u^3}{2}+mu^2+ (\frac{\epsilon
\tilde{\mathcal{L}} \tilde{K}}{2})u + (E^2 \tilde{\mathcal{L}} +
\epsilon \tilde{\mathcal{L}} m -\frac{1}{\tilde{l}^2}) +
(\frac{-\epsilon\tilde{\mathcal{L}}}{\tilde{l}^2})\frac{1}{u^2}.
\end{equation}

We will consider it for both particle and light ray as following.

\section*{3.1) Null geodesics}

For $\epsilon =0$ Eq.(\ref{du}) changes to
\begin{equation} \label{duu}
(\frac{du}{d\phi})^2=\frac{\tilde{K}u^3}{2}+mu^2 + (E^2
\tilde{\mathcal{L}} - \frac{1}{\tilde{l}^2}) =P_3 (u)=\sum_{i=0}^3
a_i u^i,
\end{equation}
which is of elliptic type. Another substitution
$u=\frac{1}{a_3}(4y-\frac{a_2}{3})=
\frac{2}{\tilde{K}}(4y-\frac{m}{3})$ transforms Eq.(\ref{duu}) into
Weierstrass form as below
\begin{equation} \label{Py}
(\frac{dy}{d\phi})^2=4y^3-\alpha y -\gamma =P_3(y),
\end{equation}
in which
\begin{equation}
\alpha =\frac{a_2 ^2}{12} - \frac{a_1
a_3}{4}=\frac{m^2}{12}\,\,\,\,,\,\,\,\,\gamma =\frac{a_1 a_2
a_3}{48} - \frac{a_0 a_3^2}{16} - \frac{a_2^3}{216} = - \frac{(E^2
\tilde{\mathcal{L}} \tilde{l}^2 -1) \tilde{K}^2}{64 \tilde{l}^2} -
\frac{m^3}{216},
\end{equation}
are Weierstrass constants. The Eq.(\ref{Py}) is of elliptic type and
is solved by the Weierstrass function
\cite{Hackmann:2008zz}, \cite{Soroushfar:2015wqa}
\begin{equation}
y(\phi)=\wp (\phi-\phi _{in} ;\alpha ,\gamma),
\end{equation}
which here we have $\phi _{in} =\phi _0 + \int _{y_0} ^\infty
\frac{dy}{\sqrt{4y^3 - \alpha y - \gamma}}$ and $y _0 =
\frac{1}{4}(\frac{a_3}{\tilde{r} _0 } + \frac{a_2}{3}) =
\frac{\tilde{K}}{8\tilde{r}_0} + \frac{m}{12}$ depends only on the
initial values $\phi_0$ and $\tilde{r}_0$. As a result, the
analytical solution of Eq.(\ref{tilde}) is
\begin{equation}
\tilde{r} (\phi) = \frac{a_3}{4\wp (\phi - \phi _{in} ; \alpha ;
\gamma) - \frac{a_2}{3}} = \frac{\tilde{K}}{2[4\wp (\phi - \phi
_{in} ; \alpha ; \gamma) - \frac{m}{3} ]}.
\end{equation}

Using this solution we could create the examples of null geodesics
for each region of different types of orbits which are plotted in
Figs.\ref{bb} and \ref{nn}.

\section*{3.2) Timelike geodesics}

For $\epsilon =1$  Eq.(\ref{du}) changes to
\begin{equation}\label{dovom}
(u\frac{du}{d\phi})^2=\frac{\tilde{K}
u^5}{2}+mu^4+(\frac{\tilde{\mathcal{L}} \tilde{K}}{2})u^3 + (E^2
\tilde{\mathcal{L}} +\tilde{\mathcal{L}} m
-\frac{1}{\tilde{l}^2})u^2 - \frac{\tilde{\mathcal{L}}}{\tilde{l}^2}
=P_5 (u)= \sum_{i=1} ^5 a_i u^i,
\end{equation}
which is a polynomial of degree $5$ with an analytical solution as
below \cite{Hackmann:2008zz, Enolski:2010if, Soroushfar:2015wqa}
\begin{equation}\label{udu}
u(\phi)=-\frac{\sigma _1}{\sigma _2}(\phi _\sigma),
\end{equation}
where $ \sigma _i $ is the i-th derivative of the Kleinian sigma
function in two variables
\begin{equation}
\sigma(z)=C e^{-\frac{1}{2}z^t \eta \omega^{-1} z} \theta
[g,h]((2\omega) ^{-1} z ; \tau).
\end{equation}

We have some parameters here: the symmetric Riemann matrix
$\tau=\omega^{-1}\acute{\omega}$, the Riemann theta-function
$\theta[g,h]$, which is written as
\begin{equation}
\theta[g;h](z;\tau)=\sum_{m\in \mathbb{Z}^g} e^{i\pi (m+g)^t(\tau
(m+g)+2z+2h)},
\end{equation}
the period-matrix $(2\omega , 2\acute{\omega})$, the period-matrix
of the second type $(2\eta , 2\acute{\eta})$, and $C$ is a constant
that can be given explicitly. Note that $z$ is a zero of the
Kleinian sigma function if and only if $(2\omega)^{-1} z$ is a zero
of the theta-function $\theta [g,h]$.

With Eq.(\ref{udu}) the solution for $\tilde{r}$ is
\begin{equation}
\tilde{r}=-\frac{\sigma _2}{\sigma _1}(\phi _\sigma) .
\end{equation}

This solution of
$\tilde{r}$ is the analytical solution of the equation of motion for
massive particle. Different types of orbits for each region of this solution
are illustrated in Figs.\ref{cc}--\ref{aa}.

\subsection{Orbits} \label{orbit}

\begin{figure}[ht]
\centerline{\includegraphics[width=7.25cm]{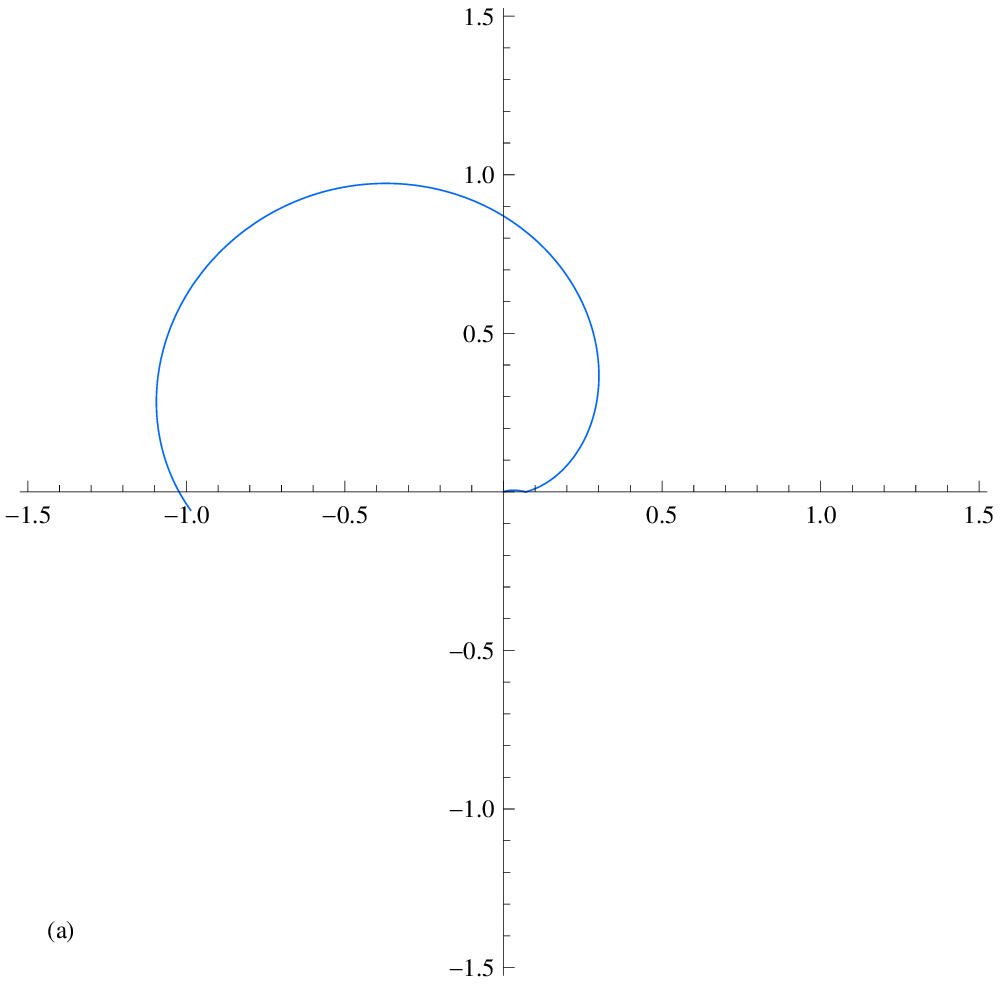}}
\centerline{\includegraphics[width=7.25cm]{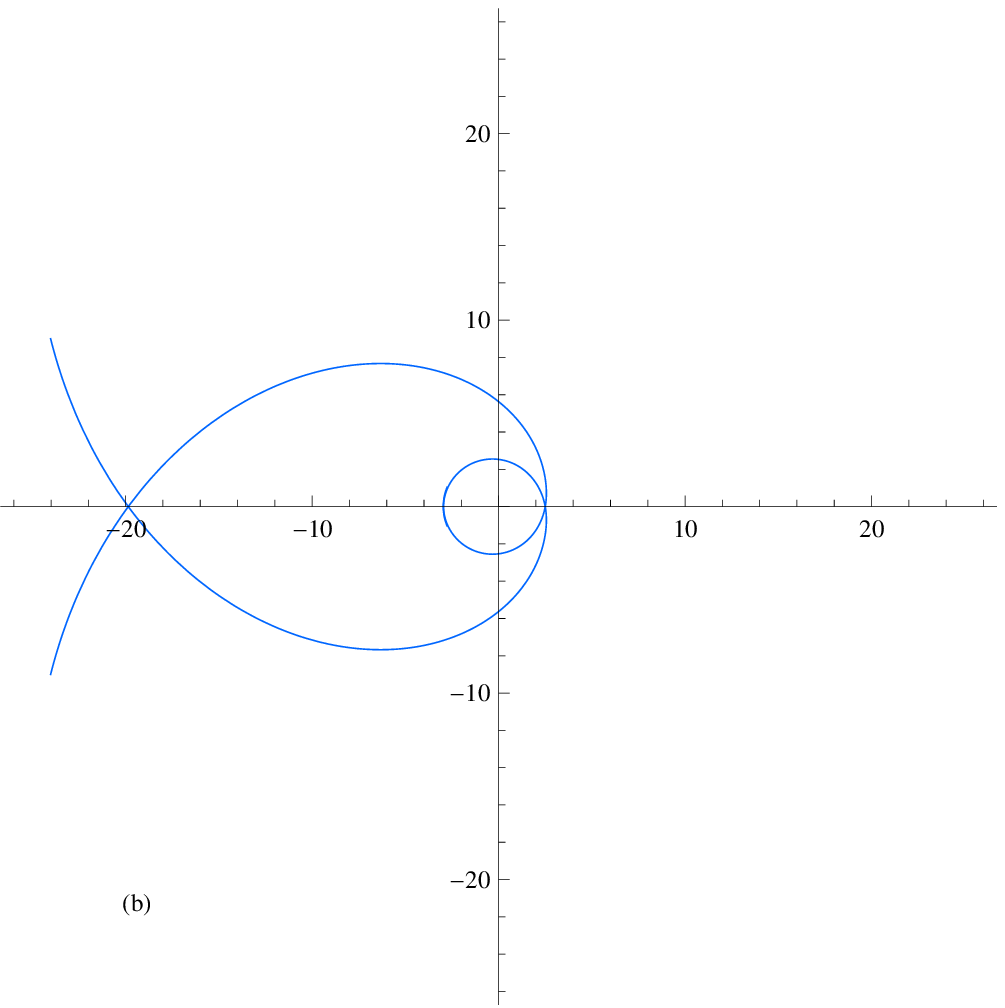}}
\caption{\label{bb}\small Null geodesic, Region I: (a)
corresponding Terminating Bound Orbit with $E^2=0.3$,
$\mathcal{L}=0.1$ , (b) corresponding Flyby Orbit with $E^2=0.9$,
$\mathcal{L}=0.1$. }
\end{figure}

\begin{figure}[ht]
\centerline{\includegraphics[width=7.25cm]{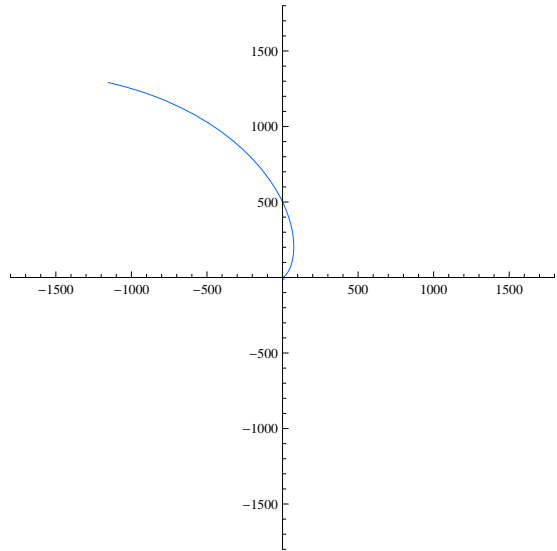}}
\caption{\label{nn}\small Null geodesic, Region III: corresponding
Terminating Escape Orbit with $E^2=3$, $\mathcal{L}=0.3$. }
\end{figure}

\begin{figure}[ht]
\centerline{\includegraphics[width=7.25cm]{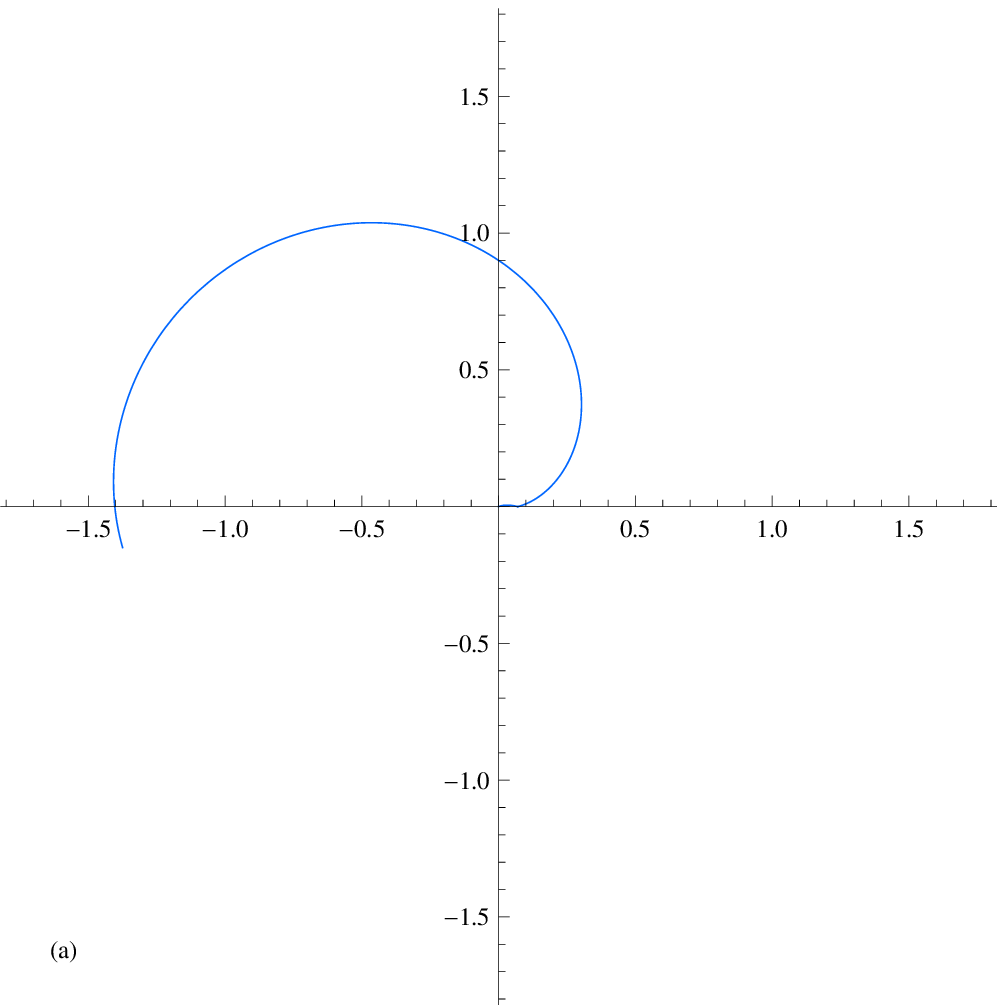}}
\centerline{\includegraphics[width=7.25cm]{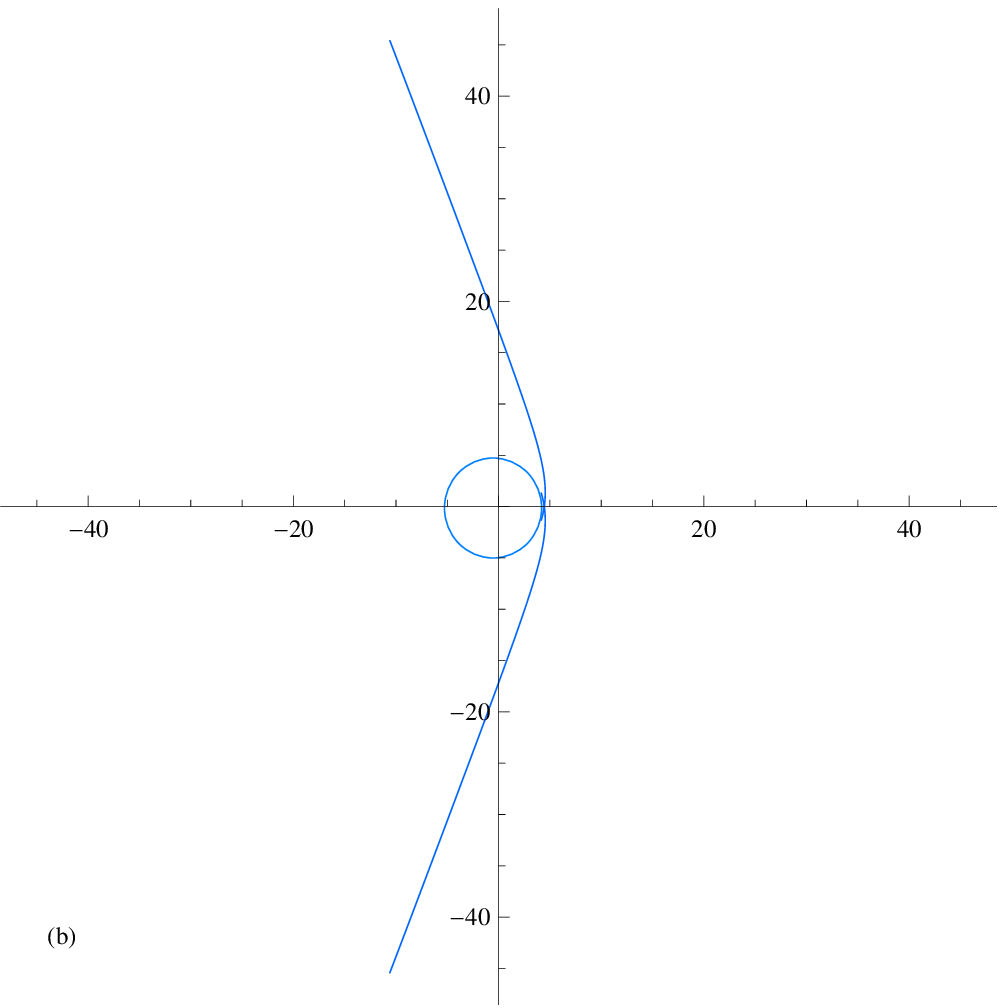}}
\caption{\label{cc}\small Timelike geodesic, Region I: (a)
corresponding Terminating Bound Orbit with $E^2=1.2$,
$\mathcal{L}=0.11$ , (b) corresponding Flyby Orbit with $E^2=1.6$,
$\mathcal{L}=0.05$. }
\end{figure}

\begin{figure}[ht]
\centerline{\includegraphics[width=7.25cm]{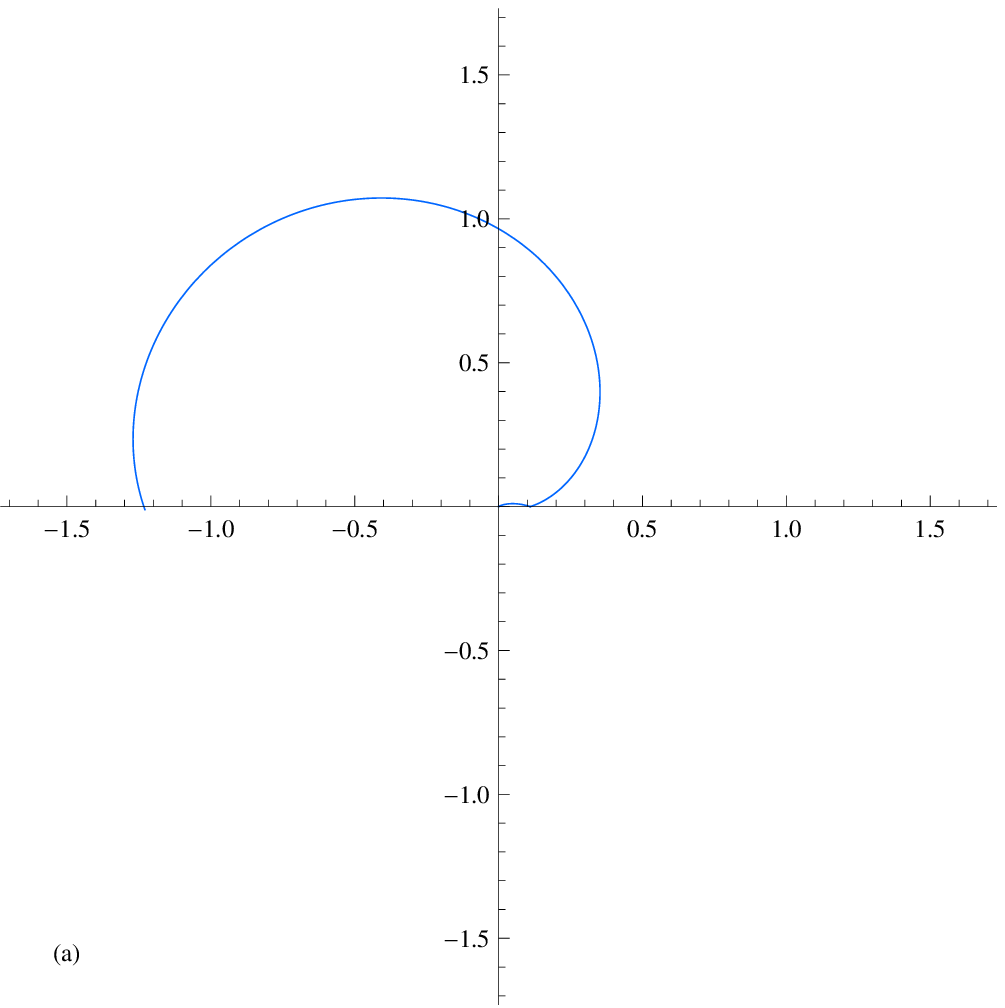}}
\centerline{\includegraphics[width=7.25cm]{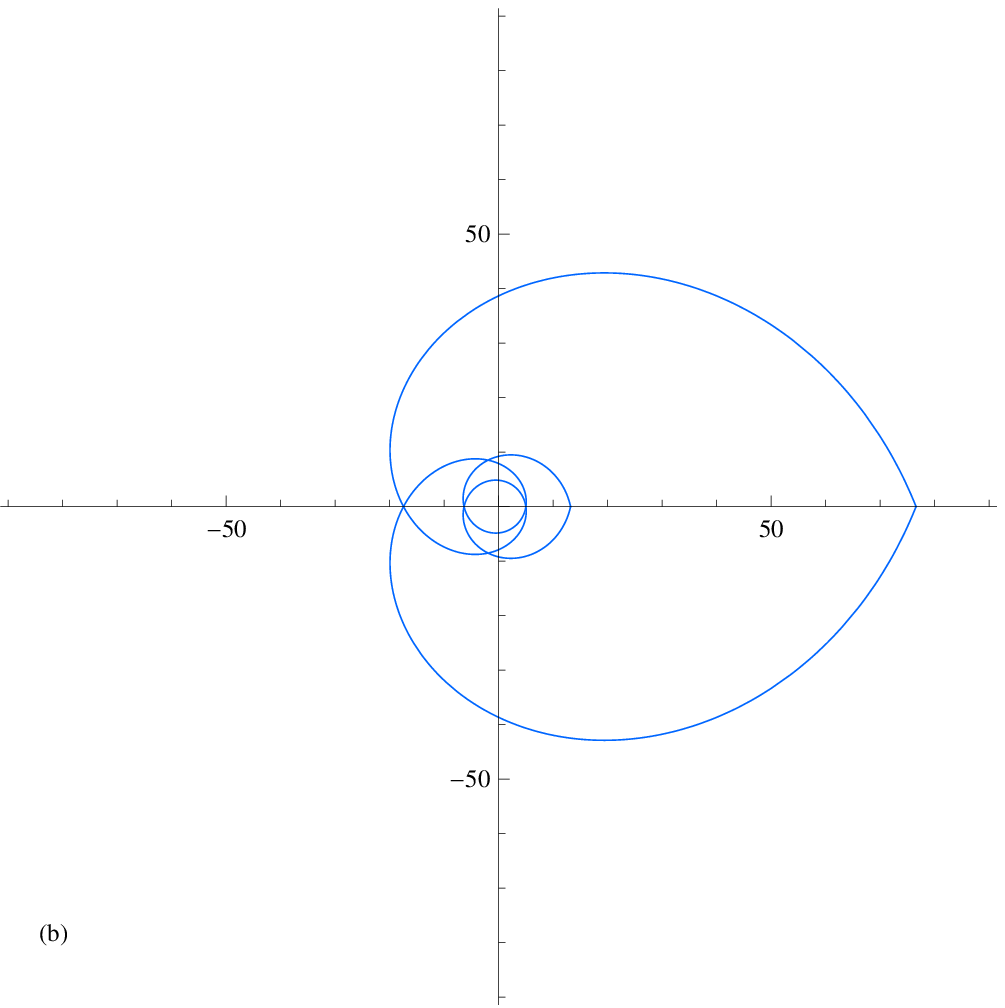}}
\centerline{\includegraphics[width=7.25cm]{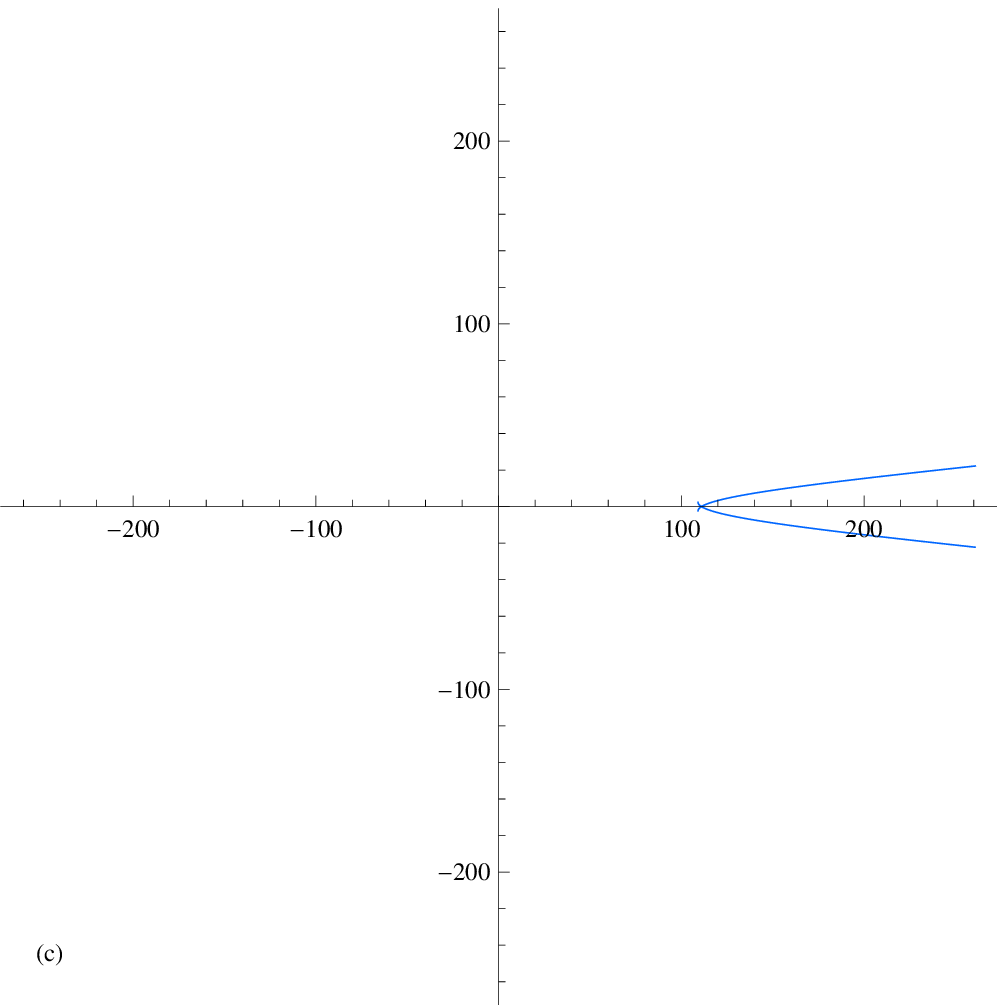}}
\caption{\label{mm}\small Timelike geodesic, Region II: (a)
corresponding Terminating Bound Orbit with $E^2=0.965$,
$\mathcal{L}=0.11$ , (b) corresponding Bound Orbit with $E^2=0.95$,
$\mathcal{L}=0.17$, (c) corresponding Flyby Orbit with $E^2=0.95$,
$\mathcal{L}=0.17$. }
\end{figure}

\begin{figure}[ht]
\centerline{\includegraphics[width=7.25cm]{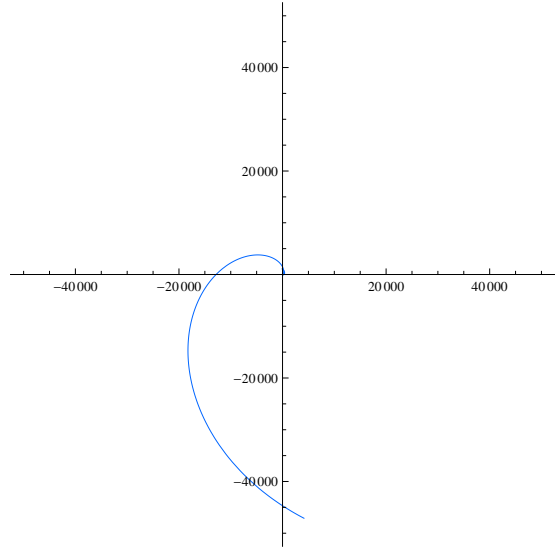}}
\caption{\label{aa}\small Timelike geodesic, Region III:
Terminating Escape Orbit with $E^2=2.2$, $\mathcal{L}=0.15$. }
\end{figure}

 In region I, as we expressed before, there are two
kinds of orbits ((TBO: $r$ starts in $(0,r_a]$ for $0<r_a<\infty$
and falls into the singulariry at $r=0$),(FO: $r$ starts from
$\infty$, then approaches a periapsis $r=r_p$ and then goes back to
$\infty$)) with $E^2=1.2$ and $\mathcal{L}=0.11$. In region II, we have
three orbits ((TBO),(FO),(BO: $r$ oscillates between two boundary
values $r_p \leq r\leq r_a$ with $0< r_p<r_a <\infty$)) with
$E^2=0.95$ and $\mathcal{L}=0.17$. Region III has just one kind of
orbit (TEO: $r$ comes from $\infty$ and falls into the singularity
at $r=0$) with $E^2=1.2$ and $\mathcal{L}=0.15$. With the help of analytical solutions,
parameter diagrams Figs.~\ref{massivespacetime},
\ref{masslessspacetime} and effective
potentials Fig.~\ref{potentials}, various orbits for these three regions considering,
$\Lambda = -\frac{1}{l^{2}}=\frac{1}{3}(10^{-5})$ and $q=1.25$, are presented
in Figs.~\ref{bb}--\ref{aa}.

\clearpage

\section{Conclusion} \label{con}

In this paper, considering a three dimensional charged BTZ black
hole, we studied the motion of particles (massive) and light rays
(massless). For this purpose, at first we found equations of motion 
(geodesic equations), then using effective potential and
solving geodesic equations in terms of Weierstrass elliptic
function and Kleinian sigma hyper-elliptic function, we classified
the complete set of orbit types. We also demonstrated that for both
timelike and null geodesics there are different regions where test
particles can move in. These regions and
possible kinds of motion are illustrated in Figs.
\ref{massivespacetime}--\ref{aa}. For timelike geodesics TBO,
BO, FO and TEO are possible and for null geodesics TBO, FO and TEO
are possible.

These results and obtained figures can be used to have an intuition
about the properties of the orbits such as light deflection,
periastron shift and so on. The higher dimension and rotating
version of this spacetime could be studied in future.

\bibliographystyle{amsplain}

\begin{thebibliography}{5}

\bibitem{Singh:2014gva}
  D.~V.~Singh and S.~Siwach,
  J.\ Phys.\ Conf.\ Ser.\  {\bf 481}, 012014 (2014).

\bibitem{Banados:1992wn}
  M.~Banados, C.~Teitelboim and J.~Zanelli,
  Phys.\ Rev.\ Lett.\  {\bf 69}, 1849 (1992)
  [hep-th/9204099].
  
\bibitem{Ross:1992ba}
  S.~F.~Ross and R.~B.~Mann,
  Phys.\ Rev.\ D {\bf 47}, 3319 (1993)
  [hep-th/9208036].


\bibitem{Horowitz:1993jc}
  G.~T.~Horowitz and D.~L.~Welch,
  Phys.\ Rev.\ Lett.\  {\bf 71}, 328 (1993)
  [hep-th/9302126].

\bibitem{Banados:1992gq}
  M.~Banados, M.~Henneaux, C.~Teitelboim and J.~Zanelli,
  Phys.\ Rev.\ D {\bf 48}, 1506 (1993)
  [Phys.\ Rev.\ D {\bf 88}, no. 6, 069902 (2013)]
  [gr-qc/9302012].

\bibitem{Martinez:1999qi}
  C.~Martinez, C.~Teitelboim and J.~Zanelli,
  Phys.\ Rev.\ D {\bf 61}, 104013 (2000)
  [hep-th/9912259].


\bibitem{Carlip:1995qv}
  S.~Carlip,
  Class.\ Quant.\ Grav.\  {\bf 12}, 2853 (1995)
  [gr-qc/9506079].

\bibitem{Clement:1995zt}
  G.~Clement,
  Phys.\ Lett.\ B {\bf 367}, 70 (1996)
  [gr-qc/9510025].


\bibitem{Sfetsos:1997xs}
  K.~Sfetsos and K.~Skenderis,
  Nucl.\ Phys.\ B {\bf 517}, 179 (1998)
  [hep-th/9711138].

\bibitem{Hyun:1997jv}
  S.~Hyun,
  J.\ Korean Phys.\ Soc.\  {\bf 33}, S532 (1998)
  [hep-th/9704005].

\bibitem{Carlip:1994gy}
  S.~Carlip,
  Phys.\ Rev.\ D {\bf 51}, 632 (1995)
  [gr-qc/9409052].

\bibitem{Carlip:1996yb}
  S.~Carlip,
  Phys.\ Rev.\ D {\bf 55}, 878 (1997)
  [gr-qc/9606043].

\bibitem{Strominger:1996sh}
  A.~Strominger and C.~Vafa,
  Phys.\ Lett.\ B {\bf 379}, 99 (1996)
  [hep-th/9601029].


  \bibitem{Y.Hagihara:1931}
  Y.Hagihara. \textit{Theory of relativistic trajectories in a gravitational field of Schwarzschild} .( Japan. J. Astron.Geophys., 8:67, 1931).

\bibitem{Hackmann:2008zz}
  E.~Hackmann and C.~Lammerzahl,
  Phys.\ Rev.\ D {\bf 78}, 024035 (2008)
  [arXiv:1505.07973 [gr-qc]].

\bibitem{Hackmann:2008tu}
  E.~Hackmann, V.~Kagramanova, J.~Kunz and C.~Lammerzahl,
  Phys.\ Rev.\ D {\bf 78}, 124018 (2008); Phys.\ Rev.\ D {\bf 79}, 029901 (E) (2009) 
  [arXiv:0812.2428 [gr-qc]].

\bibitem{Kerr:1963ud}
  R.~P.~Kerr,
  Phys.\ Rev.\ Lett.\  {\bf 11}, 237 (1963).

\bibitem{Hackmann:2010zz}
  E.~Hackmann, C.~Lammerzahl, V.~Kagramanova and J.~Kunz,
  Phys.\ Rev.\ D {\bf 81}, 044020 (2010)
  [arXiv:1009.6117 [gr-qc]].

\bibitem{Soroushfar:2015wqa}
  S.~Soroushfar, R.~Saffari, J.~Kunz and C.~Lämmerzahl,
  Phys.\ Rev.\ D {\bf 92}, 044010 (2015)
  [arXiv:1504.07854 [gr-qc]].


\bibitem{Ashtekar:2002qc} 
  A.~Ashtekar, J.~Wisniewski and O.~Dreyer,
  Adv.\ Theor.\ Math.\ Phys.\  {\bf 6}, 507 (2003)
  [gr-qc/0206024].
  
\bibitem{Sa:1995vs} 
  P.~M.~Sa, A.~Kleber and J.~P.~S.~Lemos,
  Class.\ Quant.\ Grav.\  {\bf 13}, 125 (1996)
  [hep-th/9503089].

\bibitem{Hassaine:2008pw}
  M.~Hassaine and C.~Martinez,
  Class.\ Quant.\ Grav.\  {\bf 25}, 195023 (2008)
  [arXiv:0803.2946 [hep-th]].

\bibitem{Maeda:2008ha}
  H.~Maeda, M.~Hassaine and C.~Martinez,
  Phys.\ Rev.\ D {\bf 79}, 044012 (2009)
  [arXiv:0812.2038 [gr-qc]].
\bibitem{Hendi:2009zzc}
  S.~H.~Hendi,
  Phys.\ Lett.\ B {\bf 678}, 438 (2009)
  [arXiv:1007.2476 [hep-th]].

\bibitem{Hendi:2010px}
  S.~H.~Hendi,
  Eur.\ Phys.\ J.\ C {\bf 71}, 1551 (2011)
  [arXiv:1007.2704 [gr-qc]].

\bibitem{Hendi:2014mba}
  S.~H.~Hendi, B.~Eslam Panah and R.~Saffari,
  Int.\ J.\ Mod.\ Phys.\ D {\bf 23}, 1450088 (2014)
  [arXiv:1408.5570 [hep-th]].

\bibitem{Hartmann:2010rr} 
  B.~Hartmann and P.~Sirimachan,
  JHEP {\bf 1008}, 110 (2010)
  doi:10.1007/JHEP08(2010)110
  [arXiv:1007.0863 [gr-qc]].

\bibitem{Enolski:2010if}
  V.~Z.~Enolski, E.~Hackmann, V.~Kagramanova, J.~Kunz and C.~Lammerzahl,
  J.\ Geom.\ Phys.\  {\bf 61}, 899 (2011)
  [arXiv:1011.6459 [gr-qc]].

\end{thebibliography}

\end{document}